\documentstyle[11pt]{article}

\newcommand{\lsim}{
 \mathrel{\setbox0=\hbox{$<$}\raise0.6ex\copy0\kern-\wd0
 \lower0.65ex\hbox{$\sim$}}}

\newcommand{\gsim}{
 \mathrel{\setbox0=\hbox{$>$}\raise0.6ex\copy0\kern-\wd0
 \lower0.65ex\hbox{$\sim$}}}

\begin{document}

\begin{titlepage}
\vspace*{-2cm}
\begin{flushright}
\bf
ADP-95-25/T180\\
TUM/T39-95-8
\end{flushright}

\bigskip

\begin{center}
  {\large\bf $J/\psi$ production as a probe of charge symmetry
    violations \protect\\ and nuclear corrections in parton
distributions$^{*}$}

  \vspace{2.cm}

  {\large G. Piller$^{1}$ and
    A. W. Thomas$^{2}$ }
  \bigskip

  $^{1}$Physik Department, Technische
  Universit\"at M\"unchen,\\ D-85747
  Garching, Germany\\
  $^{2}$Department of Physics and
  Mathematical Physics, University of
  Adelaide,\\ S.A. 5005, Australia

  \vspace{5.cm}

  {\bf Abstract}
\end{center}

 We investigate the sensitivity of $J/\Psi$ production in proton-neutron
 and proton-deuteron
 collisions to charge symmetry violations in the parton distributions as
 well as to nuclear corrections. It is found that the effects of charge
 symmetry violations in the quark distributions are confined to large
 $x_F$ and difficult to measure in experiments currently being planned.
 However, from proton-deuteron experiments it should be
 possible to isolate nuclear corrections to the gluon distribution.

\vspace*{1.cm}

{\sl \centerline {To be published in Z. Phys. C.}}

\vspace*{2.cm}

\noindent $^{*}$) Work supported in part
by grants from BMFT  and ARC.

\end{titlepage}

\section{Introduction}

 The production of heavy quarkonia has
 been studied quite extensively in deep-inelastic scattering
 (see e.g. ref.\cite {Allaea91})  and in hadron-hadron
 collisions (see e.g. ref.{\cite{Kowiea94,DualModel}).
 In the latter, at sufficiently high energy, the reaction
 proceeds through the annihilation of either a quark or gluon in one
 hadron with an antiquark or gluon in the other hadron. In principle one
 therefore has a probe of the quark and gluon distributions. For the
 quarks this process merely supplements the information obtained in
 Drell-Yan and deep-inelastic scattering, but for the gluons this is
 one of our few direct probes.

 There have recently been suggestions that there may be a somewhat
 larger violation of charge symmetry in the minority valence quark
 distribution in the nucleon than one might naively have expected. At
 intermediate $x$ this effect could be as large as 5\%
 \cite{Sather92,RoThLo94}. It is important to try to test these ideas in
 Drell-Yan reactions \cite{Londea94}.
 Here we examine whether such an effect could induce a significant
 forward-backward asymmetry in the production of $J/\Psi$ in proton-neutron
 collisions. As there has been no explicit calculation of a
 difference in the gluon distribution of a proton and neutron
 we also comment on this possibility.

 In proton-deuteron collisions the forward-backward asymmetry of
 $J/\psi$ production is sensitive not only to charge symmetry
 violations but also to nuclear modifications of the deuteron
 parton distributions. The latter could be investigated through
 experiments feasible at RHIC.

\section{$J/\psi$ production in proton-neutron collisions}

The production of charm-anticharm quark pairs
proceeds in leading order in the strong coupling via gluon-gluon
fusion or quark-antiquark annihilation.
The corresponding cross sections for these subprocesses are \cite{Fritsch77}:
\begin{eqnarray}
\hat \sigma_{g g} (c\bar c,M^2) &=&
\frac{\pi \alpha_s^2}{3 M^2} \left[\left( 1+4
\frac{m_c^2}{M^2} + \frac{m_c^4}{M^4} \right) ln\left(\frac{1+\lambda}
{1-\lambda}\right)- \right.\nonumber\\
&&\left. \hspace{1.5cm} \frac{\lambda}{4}
\left( 7+31\frac{m_c^2}{M^2}\right)\right],\\
\hat \sigma_{q\bar q}(c\bar c,M^2) &=&
\frac{8 \pi \alpha_s^2}{27 M^4} \left( M^2 + 2 m_c^2\right)
\lambda,
\end{eqnarray}
where  $\lambda=\sqrt{1-4 {m_c^2}/{M^2}}$.  The mass of the charm quark
and the produced $c\bar c$ pair is denoted by $m_c$ and $M$ respectively.
The strong coupling constant $\alpha_s$ is calculated at $M^2$
using a QCD scale $\Lambda = 0.177 \,GeV$.
Multiplying the cross sections of the QCD subprocesses
with the parton distributions of beam and target, $f^b_i$ and $f^t_j$, yields
the  $c\bar c$ production cross section
\begin{equation}
\frac{d^2\sigma(c\bar c) }{d x_F d M^2} = \sum_{i,j} f^b_i(x_b) f^t_j(x_t)
\frac{\hat \sigma_{ij}(c\bar c,M^2)}{s\,\sqrt{x_F^2 + \frac{4M^2}{s}}}.
\end{equation}
In the center of mass frame Feynman $x_F$ is defined as the
fraction of beam momentum carried by the produced $c\bar c$ pair
and $s$ stands for the squared  center of mass energy.
The light-cone momentum fractions of the active beam and
target parton are
\begin{eqnarray}
x_b = \frac{1}{2} \sqrt{x_F^2 + \frac{4 M^2}{s}} + \frac{1}{2} x_F,\\
x_t = \frac{1}{2} \sqrt{x_F^2 + \frac{4 M^2}{s}} - \frac{1}{2} x_F.
\end{eqnarray}

The $J/\psi$ production cross section is proportional to the
charm production cross section, integrated over the invariant mass
of the $c\bar c$ pair \cite{Fritsch77}.
The integration limits are the thresholds for
$c\bar c$ pair production and $D\bar D$ production:
\begin{equation}
\frac{d\sigma(J/\psi)}{d x_F}  = F \int_{4 m_c^2}^{4 m_D^2} dM^2\,
\frac{d^2\sigma(c\bar c)}{d x_F d M^2}.
\end{equation}
The factor $F$ specifies the fraction of events in  which $J/\psi$ bound
states are formed.
Despite being a simple model, such a description is well
known to describe many features of heavy quark production,
including the dependence of the $J/\psi$ production cross
section on $x_F$  and the beam energy \cite{Kowiea94,DualModel}.
Since it is  of importance for our further discussion, we show
in Fig.~1 the separate contributions to the $J/\psi$ production
cross section from gluon fusion and quark-antiquark annihilation.
We take $m_c=1.5\, GeV$ and use the parton distributions
of ref.\cite{Owens91}.
At small $x_F$ gluon fusion is by
far the dominant mechanism, while for $x_F>0.6$ the $q\bar q$
annihilation takes over.

To investigate charge symmetry violation we will now
focus on $J/\psi$ production in  proton-neutron collisions.
The corresponding production cross section involves
the following combination of parton distributions
\begin{eqnarray}
\hat \sigma_{q\bar q}&& \hspace{-0.3cm}
\left[u_p(x_b) \bar u_n(x_t) + \bar u_p(x_b) u_n(x_t)\,+\right.\nonumber\\
&&\;\left.
\hspace{-0.3cm} d_p(x_b) \bar d_n(x_t) + \bar d_p(x_b) d_n(x_t)\right]
\,+\nonumber\\
\hat \sigma_{gg} && \hspace{-0.3cm} g_p(x_b) g_n(x_t)
\end{eqnarray}
Here $q_{p/n}$ are the quark distributions of the proton
and neutron respectively and $g_{p/n}$ the corresponding gluon distributions.

In an isospin rotated world $J/\psi$ production in proton-neutron
collisions becomes $J/\psi$ production in neutron-proton collisions,
i.e. the role of beam and target is interchanged.
If charge symmetry were exact, the difference between the corresponding
cross sections would vanish.
Interchanging the role of beam and target is equivalent to a sign
change in $x_F$.
Hence,  the difference of the $J/\psi$ production cross sections at
positive and negative $x_F$
\begin{equation}
\Delta\sigma_{pn}(x_F) =
\left. \frac{d\sigma(J/\psi)}{d x_F}\right|_{x_F} -
\left. \frac{d\sigma(J/\psi)}{d x_F}\right|_{-x_F}
\end{equation}
is driven  by  charge symmetry violations only.
In detail, $\Delta\sigma_{pn}$  contains  the following combination
of parton distribution functions:
\begin{center}
\begin{math}
\begin{array}{rlcllcll}
 \hat \sigma_{q\bar q}\,\frac{1}{2} \!
& \Big\{\;(\delta u(x_b) \!&-&\! \delta d(x_b))
&      (\bar d(x_t)   \!&-&\! \bar u(x_t))
&        +\\
&\quad   (\delta u(x_b)  \!&+&\! \delta d(x_b))
&      (\bar d(x_t)   \!&+&\! \bar u(x_t))
&        - \\
&\quad      (u(x_b)        \!&-&\! d(x_b))
&      (\delta \bar d(x_t) \!&-&\! \delta \bar u (x_t))
&        -\\
&\quad    (u(x_b)          \!&+&\!  d(x_b))
&      (\delta \bar d (x_t) \!&+&\! \delta \bar u(x_t) ) &+
\end{array}
\end{math}
\end{center}
\vspace*{-0.6cm}
\begin{center}
\begin{math}
\quad [q \longleftrightarrow \bar q]\Big\} \,+
\end{math}
\end{center}
\vspace*{-0.4cm}
\begin{equation}  \label{Delta_pn}
\hspace*{-3.cm}\hat\sigma_{gg} \,\phantom{\frac{1}{2}}
\Big\{\delta g(x_b) g(x_t) - g(x_b) \delta g(x_t) \Big\}
\end{equation}
We expressed the neutron distributions through the proton ones,
using the definitions $\delta d \equiv d_p - u_n,\,
\delta u \equiv  u_p - d_n$ and
$\delta g \equiv g_p - g_n$, and dropping the index ``p''.

First we will consider contributions to $\Delta\sigma_{pn}$ through
charge symmetry violations in the valence distributions. The corresponding
charge symmetry violating parts of the minority and majority
distributions, $\delta d^v = d_p^v - u_n^v$ and
$\delta u^v = u_p^v - d_n^v$ were extensively discussed in
\cite{Sather92,RoThLo94}.
We use the results of ref.\cite{RoThLo94} which were obtained
within the framework of the MIT bag model.
The magnitude of $\delta d^v$ was found to be
similar to that of $\delta u^v$.
As $d^v$ is generally much larger than $u^v$ the fractional change in
$d^v$ is much greater.
This can be easily understood, since one of the major sources of
charge symmetry violation is the mass difference of the residual
di-quark pair, when one quark of the nucleon is hit in a  deep-inelastic
scattering process.
For the minority quark distribution the residual di-quark is
$uu$ in the proton and $dd$ in the neutron. Therefore in the difference
$d_p$ - $u_n$ the up-down mass difference enters twice. On the other hand,
for the majority quark distribution, where the residual system
is $ud$, both for the proton and the neutron, there is no contribution
to charge symmetry breaking.

To calculate $\Delta\sigma_{pn}$ we also need to know the
flavor asymmetry of the quark sea, $\bar d - \bar u$,
which enters in the first term of Eq.(\ref{Delta_pn}).
We take a parameterization from ref.\cite{MeThSi91}:
$x (\bar d(x) - \bar u(x)) = A  x^{0.5} (1-x)^7$,
where the normalization $A$ is fixed through
$\int dx (\bar d(x) - \bar u(x)) = 0.15$.
Such a parameterization is in good agreement with recently
discovered violations of the Gottfried sum rule \cite{NMC94}.

We normalize $\Delta\sigma_{pn} (x_F)$ through the $J/\psi$ production
cross section in proton-proton collisions and present in
Fig.~2 results for the ratio
$R_{pn}^{J/\psi} = \frac{\Delta\sigma_{pn}}{d\sigma(J/\psi)_{pp}/dx_F}$
at a center of mass energy  $\sqrt{s} = 40\,GeV$.
We find $R_{pn}^{J/\psi} \approx - 0.02$ at large
$x_F\sim 0.6-0.7$. At smaller values of  $x_F$ (say below $0.5$) the
cross section difference $\Delta\sigma_{pn}$ vanishes.
This happens for two main reasons.
At small $x_F$ gluon fusion yields the major contribution to
$J/\psi$ production and quark-antiquark annihilation
is of little relevance.
Also $\delta d^v - \delta u^v$, which is much larger than
$\delta d^v + \delta u^v$ \cite{RoThLo94}, enters
$\Delta\sigma_{pn}$
in combination with the small flavor asymmetry, $\bar d - \bar u$.

Charge symmetry breaking in the sea quark distributions has
not been calculated yet. Nevertheless its influence on
$\Delta\sigma_{pn}$ can be estimated.
In the MIT bag model the sea quark distributions are dominated by
contributions characterized through
residual four-quark states, after one quark is hit in a
deep-inelastic scattering process \cite{ScSiTh91}.
By analogy with  the valence quark case let us assume
that a major source of charge symmetry breaking is
the mass difference of the residual spectator states.
Scattering on an antiquark which carries the same flavor as
the majority quarks leaves a $uuud$ or $dddu$
residual four-quark state in the case of a proton or neutron
target respectively.
In case of an antiquark of minority flavor a $uudd$ residual
state occurs in both cases.
Since the up-down mass difference enters twice in the
first case but is absent in the second, it seems reasonable to
assume $\delta \bar u \gg \delta \bar d$.
In the following we will neglect $\delta \bar d$.

In Fig.~2 we also show  $R_{pn}^{J/\psi}$
for various values of $\delta \bar u$ between $0.01\bar u$ and $0.10\bar u$.
Qualitatively we find minor changes to our former result where
only charge symmetry breaking in the valence distributions was taken
into account.
We may therefore conclude that charge symmetry breaking
in the quark distributions contributes to the difference of
$J/\psi$ production in the forward and backward direction only at
large values of $x_F$ ($x_F\gsim 0.6$). Unfortunately this
kinematic region is not accessible at the moment.
In current measurements, using the neutron beam facility at FNAL, one is
restricted to the region $x_F > -0.1$ \cite{Moss}.

However the news is not all bad. Since charge symmetry
violations in quark distributions do not contribute
to $R_{pn}^{J/\psi}$ at small $x_F$, it is an
ideal place to look for charge
symmetry violations in the gluon distributions.
As we can see from Eq.(\ref{Delta_pn}), a (say) $1\%$
charge symmetry violation in the gluon distribution,
$\delta g \sim 0.01 g$, can lead to $|R^{J/\psi}_{pn}|\lsim 0.02$.
No predictions exist up to now for charge symmetry violations
in gluon distributions.
Since a major part of  the glue in
hadrons can be viewed as being radiatively generated from
quarks, charge symmetry violations
in the quark distributions may in principle induce similar effects
in the gluon distributions.
However, since for charge symmetry violation in the radiatively
generated  glue the small combination $\delta d^v + \delta u^v$
is relevant, a large signal is not expected. If a large signal were seen
it could only be attributed to charge symmetry
violation in the non-perturbative glue, which would certainly be
a surprise.

\section{$J/\psi$ production in proton-deuteron collisions}

In high energy processes deuterons are often used as a
convenient source of neutrons.
Furthermore, in the near future proton-deutron collisions will be
carried out at RHIC at center of mass energies
between $50$ and $375 \,GeV$ and $x_F> -0.5$ \cite{Moss}.
Therefore, at first sight it seems to be a good idea to
investigate charge symmetry violations via $J/\psi$ production
in proton-deuteron collisions.
However, as we will demonstrate below, nuclear modifications of
deuteron parton distributions, which are interesting in their own right,
are most likely to overtake the  effects resulting from charge symmetry
violations.

Let us assume that the parton distributions of the deuteron
are related to those in the proton and neutron via
\begin{eqnarray}
q_D(x) &=& (1+\epsilon_q(x)) \,(q_p(x) + q_n(x)), \quad q = u,d,\\
g_D(x) &=& (1+\epsilon_g(x)) \,(g_p(x) + g_n(x)).
\end{eqnarray}
In proton-deuteron collisions the difference $\Delta\sigma_{pD}(x_F)$
of the $J/\psi$ production cross sections  measured in the forward and
backward direction, involves the parton distribution functions:
\begin{center}
\begin{math}
\begin{array}{rlllllcll}
 \hat \sigma_{q\bar q}\, \Big\{
& \hspace{-0.3cm}\frac{1}{2} \left( 1 + \epsilon_q(x_b)\right) \big[
& \hspace{-0.3cm}     (\delta u(x_b) \!&-&\! \delta d(x_b))
& \hspace{-0.3cm}     (\bar d(x_t)   \!&-&\! \bar u(x_t))
&        +\\
&&\hspace{-0.3cm}     (\delta u(x_b)  \!&+&\! \delta d(x_b))
& \hspace{-0.3cm}      (\bar d(x_t)   \!&+&\! \bar u(x_t)) \,\big]
&        - \\
& \hspace{-0.3cm} \frac{1}{2} \left( 1 + \epsilon_q(x_t)\right) \big[
& \hspace{-0.3cm}          (u(x_b)        \!&-&\! d(x_b))
& \hspace{-0.3cm}     (\delta \bar d(x_t) \!&-&\! \delta \bar u (x_t))
&        +\\
&&\hspace{-0.3cm}     (u(x_b)          \!&+&\!  d(x_b))
& \hspace{-0.3cm}     (\delta \bar d (x_t) \!&+&\! \delta \bar u(x_t) )\big]
&+  \\
&  \hspace{-0.3cm}\left( \epsilon_q(x_t) - \epsilon_q(x_b) \right)
& \hspace{-0.3cm} (u(x_b)       \!&+&\! d(x_b))
& \hspace{-0.3cm} (\bar u(x_t) \!&+&\! \bar d(x_t))
& +
\end{array}
\end{math}
\begin{math}
[q \longleftrightarrow \bar q]\Big\} \; +
\end{math}
\end{center}
\vspace*{-0.3cm}
\begin{equation}   \label{Delta_pD}
\begin{array}{c}
\hspace*{-1.cm}\hat \sigma_{gg} \,\Big\{
\left(1 + \epsilon_g(x_b)\right)\,\delta g(x_b) \,g(x_t)\,-
\left(1 + \epsilon_g(x_t)\right)\, g(x_b) \,\delta g (x_t)\,+ \nonumber \\
\hspace{0.cm} 2 \left(\epsilon_g(x_t) - \epsilon_g(x_b)\right)
\,g(x_b) \,g(x_t)\,\Big\}
\end{array}
\end{equation}
Clearly not all contributions to $\Delta\sigma_{pD}$ arise from
charge symmetry violation. There are also terms  proportional to the
difference  of nuclear effects at $x_b$ and $x_t$.
In the following we will estimate their size and
show that they are most
likely to dominate over contributions from charge symmetry violations.

In Fig.~3 we show the light-cone momentum fractions
$x_b$ and $x_t$ at which the beam and target parton distributions
are probed  for different $x_F$ and for different center of mass
energies $\sqrt{s}$.
The invariant mass of the produced quark-antiquark pair is varied
over the range $4 m_c^2 < M^2 <  4 m_D^2$.
We observe that at large values of $s$ the dependence of
$x_{t/b}$ on $M^2$ is rather small.
While $x_t$ decreases to very small values with
increasing $x_F$, $x_b$ rises  towards $x_F$.

Let us review the present knowledge of nuclear effects in deuteron
distribution functions at light-cone momentum fractions probed
in $J/\psi$ production.
Nuclear effects  in quark distributions have been  studied a great deal
in deep-inelastic scattering processes
(see e.g. \cite{Arneodo94,AnnRev}).
At small values of Bjorken $x$ ($x<0.1$) shadowing effects
in the deuteron of about $(1-4)\%$ were predicted by many models
and also recently observed
(see \cite{NMC94,ShaDeu} and references therein).
They suggest $\epsilon_q (x) \sim (-0.01) \;\mbox{---} \; (-0.04)$ for
$x\sim 0.05 \;\mbox{---} \; 0.001$. At $ x \sim 0.1$ shadowing disappears
and
the deep-inelastic scattering cross section for nuclear targets becomes
slightly larger than the corresponding cross section for free nucleons.
Such a behavior is also expected for deuterium,
although small \cite{FrStLi90}.
It suggests $\epsilon_q (x\sim 0.1) \gsim 0$.
For  $x>0.2$ binding effects cause a decreases of
$\epsilon_q <0$, while Fermi motion leads to  a rise
at $x>0.8$ (see e.g. \cite{MeScTh94}).
Nuclear effects in gluon distributions are not so well established
up to now. However momentum and baryon number conservation suggest
that for $x<0.2$ they exhibit a behavior similar to
that of the quark distributions --
i.e. $\epsilon_g \sim \epsilon_q$ \cite{FrStLi90}.

The preceeding discussion suggests that  modifications of
the deuteron parton distributions might easily be larger than
the charge symmetry violating effects. For example, at a center of
mass energy $\sqrt{s} = 80\,GeV$ and $x_F\approx 0.1$ we have
$x_t \approx 0.02$ and $x_b\approx 0.12$.
While $x_t$ is in the shadowing domain, $x_b$ lies  in the
region where the deuteron parton distributions might be enhanced or
are at least equal to the nucleon ones.
The difference
$\Delta \epsilon_{q/g} = \epsilon_{q/g}(x_t) - \epsilon_{q/g}(x_b)$
can therefore  easily range from  $\Delta \epsilon_{q/g} \approx -0.01$ to
 $\Delta \epsilon_{q/g} \approx -0.03$

In contributions  to $\Delta\sigma_{pD}$ which are proportional
to charge symmetry violations one may neglect nuclear effects
to a good approximation.
Then nuclear effects enter $\Delta\sigma_{pD}$
via the difference $\Delta \epsilon_{q/g}$ only.
In Fig.~4 we show separately the contributions
of gluon fusion and  quark-antiquark annihilation to the ratio
$R_{pD}^{J/\psi} = \frac{\Delta\sigma_{pD}}{d\sigma(J/\psi)_{pp}/dx_F}$,
for different
$\Delta \epsilon_{q/g}$,
at a center of mass energy $\sqrt{s} = 80\,GeV$.
For the charge symmetry violation in the valence distributions we again use
the results of ref.\cite{RoThLo94},
while the charge symmetry violation in the sea and gluon distributions
are chosen to be zero.
We find that at small values of $x_F$ nuclear modifications of the
gluon distribution in deuterium dominate  $R_{pD}^{J/\psi}$ or
equivalently  $\Delta\sigma_{pD}$.
Therefore, if charge symmetry violation in the gluon distributions,
which are accessible through $R_{pn}^{J/\psi}$, are small,
nuclear modifications of the gluon distribution in deuterium
can be investigated. Figure~4
demonstrates that, at large $x_F$, nuclear effects may easily dominate
charge symmetry violation in the valence distributions.

\section{Conclusion}
We have discussed $J/\psi$ production in proton-neutron and
proton-deuteron collisions as a tool for investigating charge
symmetry breaking and nuclear modifications of parton distributions.
In proton-neutron collisions the difference of the $J/\psi$
production cross section in the forward and backward direction,
$\Delta\sigma_{pn}$, is solely due to charge symmetry violations in the
nucleon.
Charge symmetry breaking in the valence and sea quark distributions
affect  the cross section difference significantly only at large
$x_F \sim 0.6$.
At small values of $x_F\lsim 0.1$ a non vanishing result for
$\Delta\sigma_{pn}$ would be entirely due to charge symmetry
violations in the gluon distributions.
Corresponding measurements should be possible
using the neutron beam facility at FNAL.

$J/\psi$ production in the forward and backward direction
through proton-deuteron collisions will be possible at
RHIC for a wide range of center of mass energies and Feynman $x_F$.
The corresponding cross section difference
is, however, not only due to charge symmetry violations but also
to nuclear modifications of the deuteron parton distributions.
As a consequence nuclear modifications of the gluon
distribution in the deuteron could be investigated
as well.

\bigskip
\noindent
We would like to thank J. Moss for helpful discussions.


\newpage

\begin{figure}
\caption{Ratio of the quark (solid) and the gluon (dashed) contribution to
the $J/\psi$ production to their sum, for proton-proton  collisions at
$\protect{\sqrt s} = 40\,GeV$.}
\end{figure}

\begin{figure}
\caption{Ratio $R_{pn}^{J/\psi}$ at
$\protect{\sqrt{s}} = 40\,GeV$. The solid line includes charge symmetry
violations in the valence quark distributions only.
Different values for  charge symmetry violations in the sea
quark distributions are used: $\delta \bar u = 0.01 \bar u$ (dashed),
$\delta \bar u = 0.05 \bar u$ (dotted) and $\delta \bar u = 0.1 \bar u$
(dot-dashed).}
\end{figure}

\begin{figure}
\caption{$x_F$-dependence of the light-cone momentum fractions
$x_b$ (solid) and $x_t$ (dashed) for different center of mass energies.
The shaded area indicates their dependence
on the mass $4 m_c^2  < M^2 < 4 m_D^2$ of the final $c\bar c$ state.}
\end{figure}

\begin{figure}
\caption{Gluon (dashed) and quark (dotted) contributions to
the ratio $R_{pD}^{J/\psi}$ at $\protect{\sqrt{s}} = 80\,GeV$.
For the nuclear modifications of the gluon and quark distributions
in the deuteron $\Delta \epsilon_{g/q} = -0.01,-0.02$  are used.
The solid line shows $R_{pD}^{J/\psi}$  in  absence of nuclear effects.}
\end{figure}

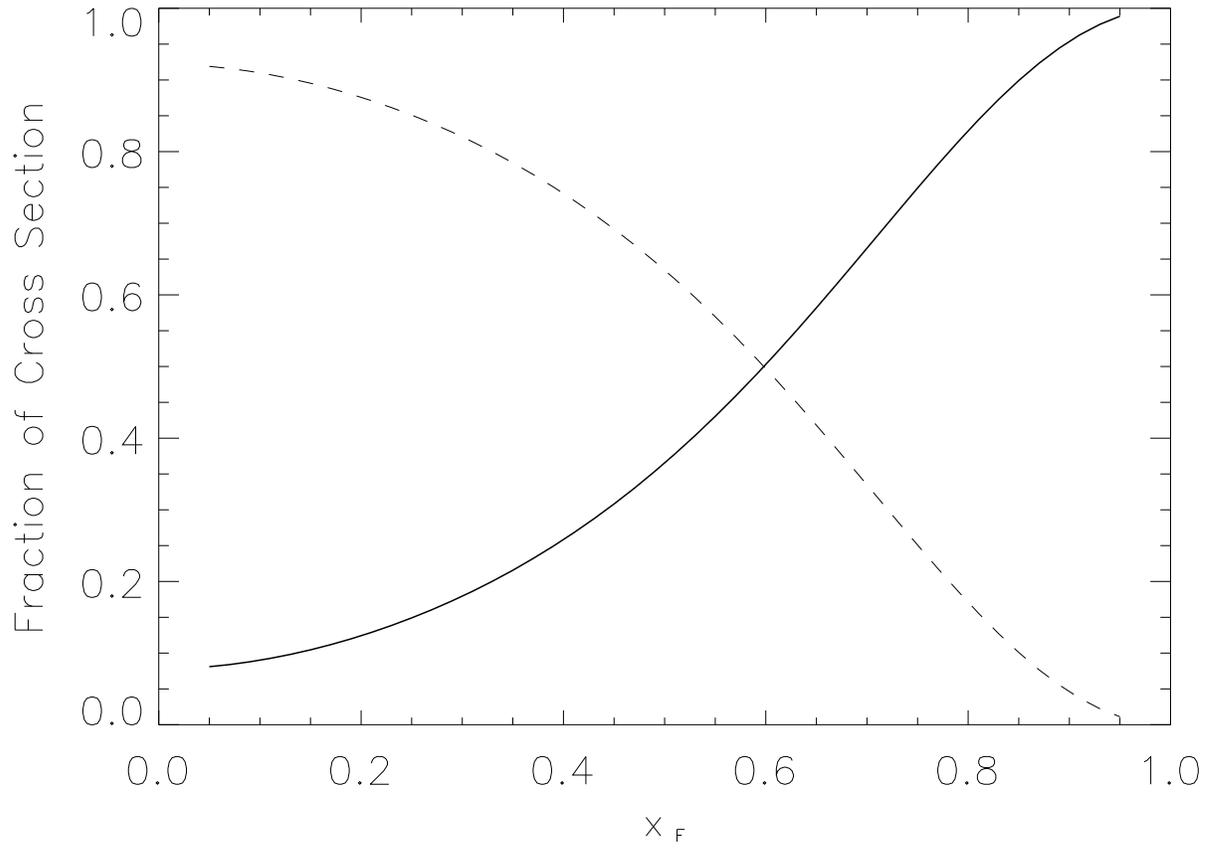
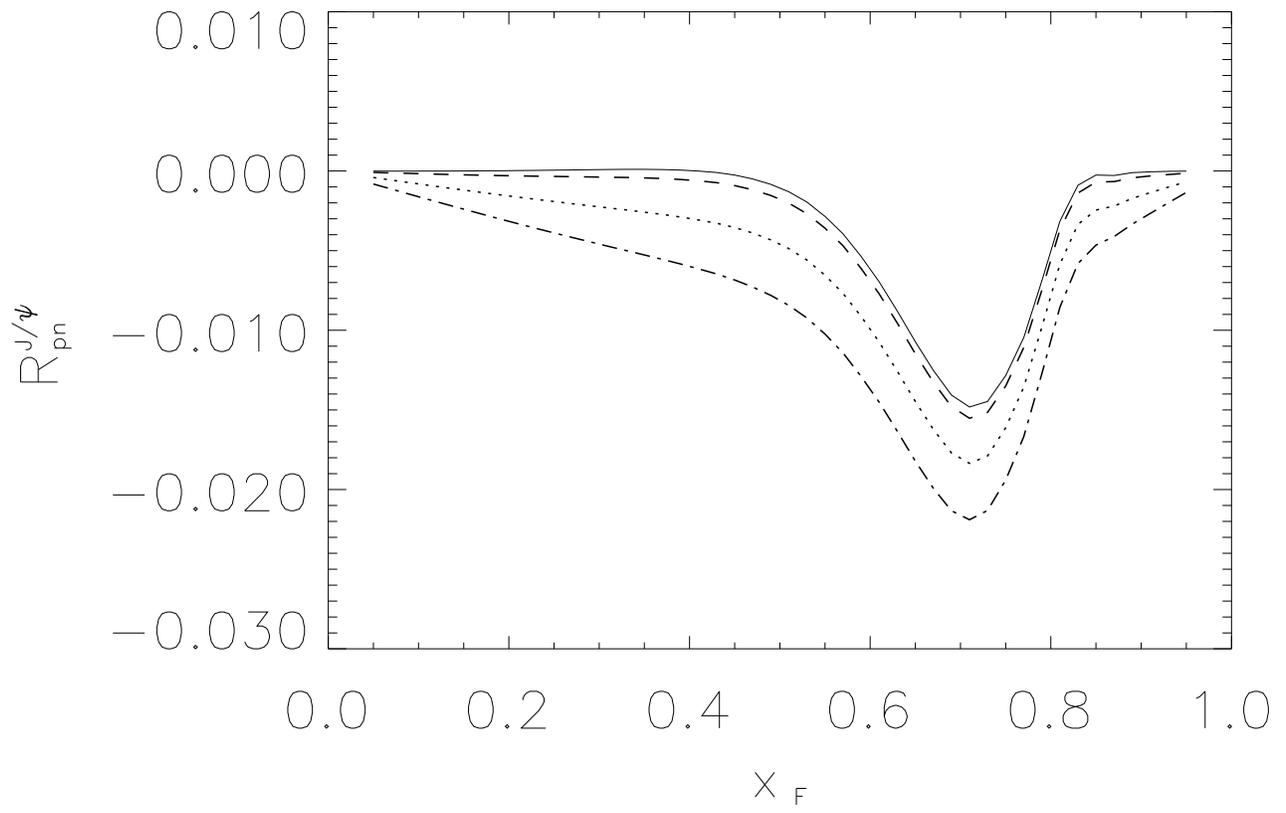
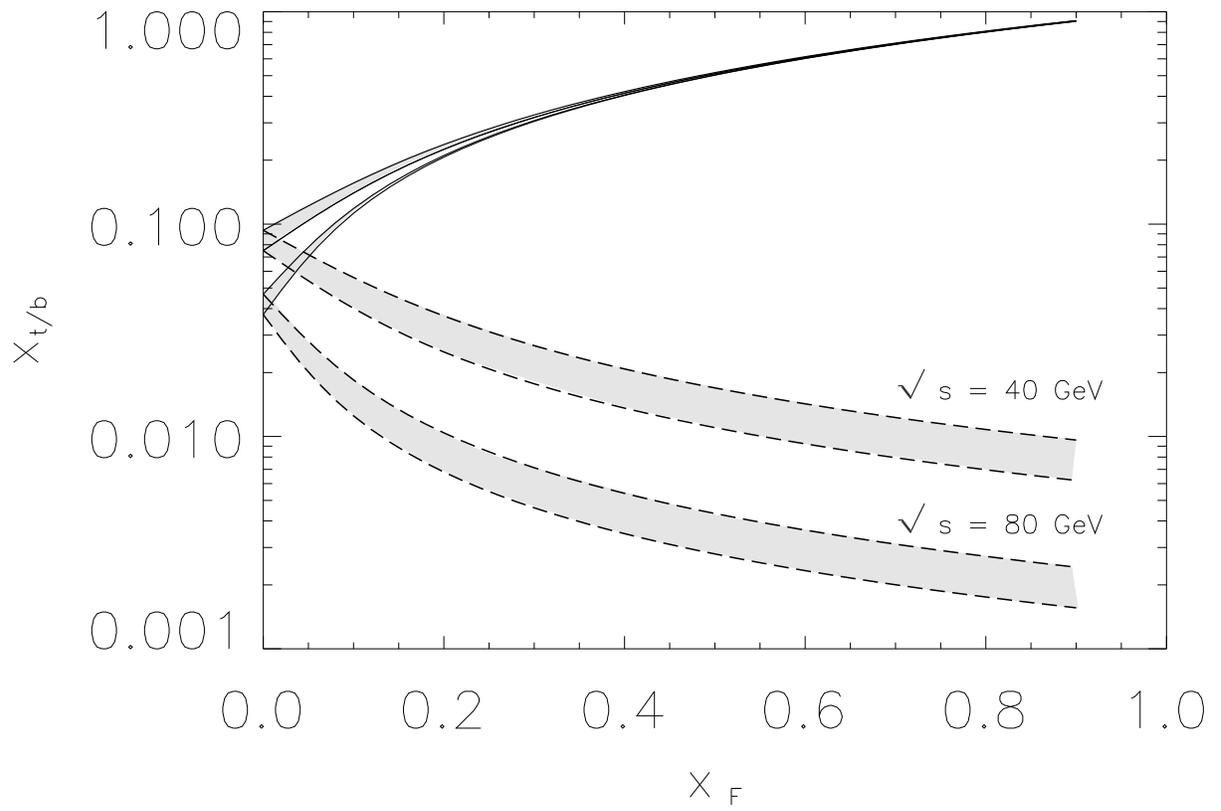
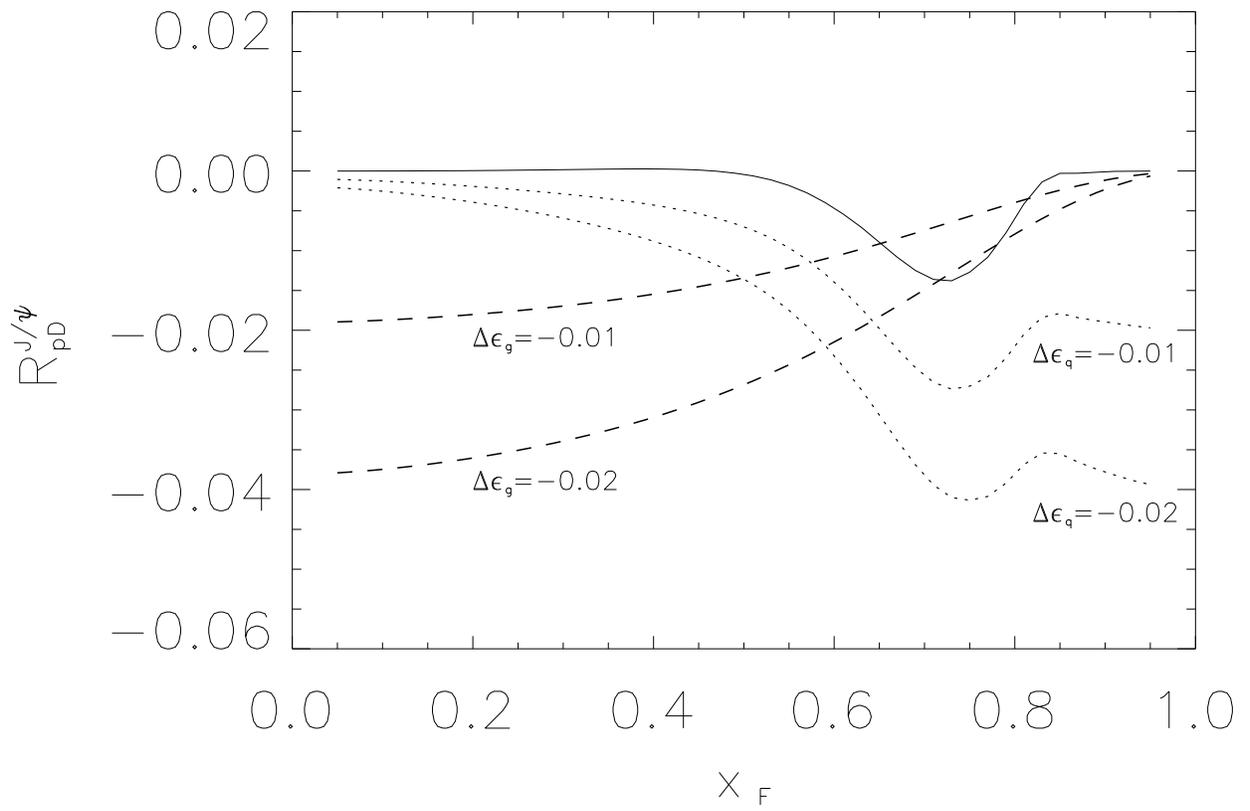
\end{document}